# Features of the gas discharge in the narrow gap micro-pattern gas detectors (MPGD) at a high level of alpha-particles background

***V.I.Razin, A.I.Reshetin***

Institute for Nuclear Research of the Russian Academy of Sciences, Moscow, Russia

## Abstract

In given article preliminary results of the research of the electron multiplication in MPGD are presented at a high level of alpha-particles background. This work has expanded borders of understanding of the streamer mode nature. It is seen as a complex from electrostatic and electromagnetic interactions which begin with appearance of the precursor in plasma state. In an inter-electrode gap the plasma oscillations occur, accompanied by longitudinal elastic waves of ionization, which can reach the cathode surface with induced negative charge. With the release of this charge due to previously established conducting channel there is a strong current pulse, accompanied by the emission due to recombination of positive and negative ions and a thin cord or streamer derive.

In the aim of the MPGD protection from the spark breakdown at a high level of the alpha-particle background the next gas composition from a buffer, cooling and electronegative components are offered: 70% He +28% $CF_4$ +2% $SF_6$.

## 1. Introduction

It is well know that narrow gap gas detectors are very convenient for a study of gas discharge phenomena because of negligible role of the feedback photon mechanism [1, 2]. Another simple factor is the practically uniform electric field between the plate electrodes in absence of wires. Nevertheless in spite of a wide area application of the MPGD at this time the question about gas amplification factor did not receive of the good explanation. There are many distinctive features of MPGD, for example, the electron multiplication at holes, the absence of photon feedback, a multistep development of the avalanches, a high gas multiple factor without sparks and etc. A study of these phenomena gives the possibility more comprehensive understanding of the nature of the streamer formation at MPGD before a spark breakdown.

In Table 1, the classification of different gas discharge modes is presented for devices with homogeneous and inhomogeneous electric fields, in which the gas pressure may be up and down than 1 bar, for example, in the case of Geiger counters. Table 1 shows that after Townsand limited proportionality mode the one from three modes may be followed in the dependence on a pressure and a configuration of electrodes:

a) Geiger mode;
b) Corona mode;
c) Streamer mode.

Table 1.

| **Gas discharge modes** | Mode of gas ionization | Townsand avalanche | Geiger mode<br>Corona mode<br>Streamer mode | Spark mode |
|---|---|---|---|---|
| **Effective current** | **pA** | **nA** | **mkA** | **mA** |

The amount of an effective current at outer circuit shows by indirect way a picture of a discharge in a range from pA to mA.

As it follows from work [2], in an avalanche with a gas gain of $10^7$ - $10^8$ the purely plasma processes, such as electron deceleration, electro-neutralization of charges, electrostatic plasma oscillations of the ion branch and other processes must be displayed, which absent in Townsend avalanches.

It is considered, that photons play the main role in the case of the transit from the avalanche to Geiger mode [3], and also it may be acceptable in the case of the corona formation. But in the case of the streamer formation the reality of this mechanism do not confirm. Indeed in the recently developed hole- type gaseous detectors the photon mechanism of electron exit from a cathode do not work because the detector surface is closed by a dielectric film. Taking into account that a free path length of the initiating photons not exceed (20 - 40) microns in the gas mixtures with cool and electronegative additions it seems that in the devices of the MICROMEGAS type the role of photons at forming a streamer is very small also. From paper [4] it is seen, that location of the streamer occurs along of a drift line of the initial avalanches including the zones with a large number of the meta-stable and excite atoms (dimers).The presence in MPGD of a gas mixture with the electronegative addition results to a large amplification factor. The positive space charge drifts to cathode. After a sometime delay the new avalanche process may have a place inside of the gap. Loeb [5] pointed out this phenomenon. He confirmed that the development of a streamer was started from a volume of the gas detector. The initial electrons for the streamer formation may be appeared from the next reactions:

$$\left.\begin{array}{lll} 1. & A^{**} + B = B^{+} + e^{-} + A & \Longrightarrow \text{meta-stable Penning effect.} \\ 2. & A^{*} + A^{*} + B = B^{+} + e^{-} + A_2 & \Longrightarrow \text{non meta-stable Penning effect.} \\ 3. & A^{*} + A = A_2^{+} + e^{-} & \Longrightarrow \text{associative ionization.} \\ 4. & A^{*} + A^{*} = A_2^{+} + e^{-} & \Longrightarrow \text{step ionization.} \end{array}\right\} (1),$$

where $A^{**}$ − the meta-stable state of A; $A^{*}$ − the excite state of A.

According to data [6] the streamer mode was not occurred in Ar or $CH_4$ separately. It needs to consider the reactions type 1, 2 as more preferably because the effective cross sections of an ionization for exciting atoms and molecules are much more than for a ground states [7].

The choice of the narrow-gap detectors as an instrument for obtaining the information about the avalanche mechanism and the streamer formation is explained by the fact that in such devices the meta-stable Penning effect is occurred.

Value of the relation E/P is averaged along the lines of force in a much higher degree than at gas detectors with wide gaps. An increase in the ratio E/P and, consequently, an expansion of the avalanche formation region are favored by the selection of the main gas components with a fairly high gas amplification factor. In this case the charge density in the avalanche rises, and, when it reaches a certain value, electric forces between charges of opposite signs begin to manifest themselves. The avalanche development dynamic under such conditions must have a qualitative different behavior because the avalanche with a gas gain of $10^7$-$10^8$ begins to transform a plasma state with the next processes:

1. Electron deceleration.
2. Electro-neutralization of charges.
3. Delay of a cathode signal.
4. Electrostatic plasma oscillations of the ion branch.
5. Precursor.

## 2. The features of the breakdown processes in MPGD under high rate conditions

The condition of a streamer appearance before spark breakdown was found by Loeb, Mik and Rather [5] in the next equation empirically:

$$\alpha(E_0) \cdot d = 20, \qquad (2)$$

where $\alpha(E_0)$ – the ionization coefficient under $E_0$ ,
d – the dimension of a gap.

A gas gain may to reach the amount of (5 - 10) $\cdot 10^7$, when the equation (2) is performed. It is considered, that electric field in the region of the streamer formation is equal or more than in the case of the initial avalanche and the streamer have the direct to cathode. It points out also that a delay time between the first avalanche and streamer formation has the definite value in dependence on the size of d. What is concerning of the precursor this phenomenon follows from the low of similarity for the gas discharge, which shows that the different modes have the same features. For example, the precursor exists as channel-leader at storm discharge or in the case of Geiger discharge as the delay pulse at the beginning of a counting plateau.

The appearance of the precursor in MPGD means that necessary conditions are created for the transition of Townsend avalanche to the plasma state which depends on the gas composition and the work voltage. This process is a more probability if at a local space the slow electrons are born in accordance with a scheme (1). The number $\alpha$ from the equation (2) reaches to a threshold value.

At a moment when the precursor is formed the ionization wave can be originated. Moving after to cathode with the velocity much more than an electron drift velocity, the front of this wave releases the induced negative charges and closes the conductivity channel between anode and cathode. Current splash is thus formed and the luminescence from recombination of the positive and negative ions and electrons is observed in the form of a cord or streamer. A time of the luminescence or deionization of a streamer is measured by milliseconds and even seconds. The probability of a streamer's origination in MPGD will be heavily dependent apparently on changes in values of both members in the expression (2), especially in the large background of alpha-particles. Coefficient $\alpha$ can be substantially reduced through the use of helium as a buffer gas and also by means of increasing the electronegative addition such as $SF_6$ up to two percent.

What is concerning of the inter-electrode distance d it should take into account the next consideration. From equality $10^6 \approx 2^{20}$ it implies that at very strong electric field, such as 100 kV/cm in MPGD-type the electron has time on a distance of 1 micron to acquire of the sufficient energy for exciting or ionization of the atoms of a buffer gas. So on the length of 20 microns it can be obtained the gas gain of order $10^6$. Taking in account the fact that MICROMEGAS cathode plane, in contrast to GEM, TGEM, RETGEM, located directly opposite the anode plane the formation of a streamer due to induced charge is more likely in the process of the breakdown. Time of the streamer development is very short: at the level of a few nanoseconds. Thus with regard to specifies of MPGD to enhance resistance to breakdown it should take the following measures:

1. To decrease the electrode gap d to values of 25-30 microns with no noticeable effect on efficiency.
2. To decrease the ionization coefficient α by means of the use of helium as a buffer gas.
3. To use as cooling gas additive $CF_4$ as the most radiation-resistant with high dielectric strength.
4. To increase the content of electronegative component $SF_6$ in the gas mixture up to 2% and more especially in the case of a high background of alpha-particles.

## 4. Conclusions

1. Research of features connected with the process of transition from Townsend limited proportionality mode to plasma and streamer modes in narrow-gap detectors type MPGD especially in the conditions of a high background of alpha-particles at LHC has paramount value with a view of struggle against the gas breakdown.
2. Absence of a photon feedback at operating with GEM, TGEM, RETGEM and presence of almost homogeneous electric field in MICROMEGAS can serve as additional simplifying factors by the consideration of a breakdown in gases.
3. One of the possible reasons for the emergence of a streamer as a precursor for spark is as following: at high density of charged particles in a narrow inter-electrode gap the plasma oscillations occur, accompanied by longitudinal elastic waves of ionization, which can reach the cathode surface having the induced negative charge. With the release of this charge previously established by the conducting channel there is a strong current pulse, accompanied by the emission due the recombination of positive and negative ions and a thin cord or streamer derive.
4. The sufficient interest and new methodical approach can be established in the research field of the breakdown process in MICROMEGAS, because when the electrode gap have a size of 25-100 microns all stages of the streamer formation should proceed in a much shorter time than 1 ns.
5. Following to the formula Loeb (2) in order to reduce the value α in the gas mixture it must be entered at least 2% of the electronegative gas such as $SF_6$. Under small sizes of a gas gap it should not affect the detection efficiency of charged particles. Also the helium should be used as a buffer gas and “cooling” addition in the form of $CF_4$ as a most radiation-resistant with high dielectric strength. Thus as a working gas mixture for MICROMEGAS it is possible to offer the next composition: 70% He +28% $CF_4$ +2% $SF_6$.